\documentstyle[12pt]{article}

\renewcommand{\theequation}{\arabic{section}.\arabic{equation}}

\newcommand{\be}{\begin{equation}}
\newcommand{\ee}{\end{equation}}
\newcommand{\bea}{\begin{eqnarray}}
\newcommand{\eea}{\end{eqnarray}}

\begin{document}

\thispagestyle{empty}
\vspace*{0.5cm}
\begin{center}
{\Large \bf On interpretations and constructions of \\
classical dynamical $r$-matrices\footnote{Based on a talk 
given by L.F. 
at the QTS2 symposium, 18-21 July 2001, Krak\'ow, Poland.}}
\end{center}

\vspace{1.5cm}

\begin{center}
L. Feh\'er and A. G\'abor\\

\bigskip

{\em 
Department of Theoretical Physics, University of Szeged \\
Tisza Lajos krt 84-86, H-6720 Szeged, Hungary \\
E-mail: lfeher@sol.cc.u-szeged.hu
}
\end{center}

\vspace{2.2cm}

\begin{abstract} 
In this note we complement recent results on the exchange
$r$-matrices appearing in the chiral WZNW model by providing a
direct, purely finite-dimensional description of the relationship 
between the monodromy dependent 2-form that enters the
chiral WZNW symplectic form and the exchange $r$-matrix 
that governs the corresponding Poisson brackets.
We also develop the  special case in which the exchange $r$-matrix 
becomes the `canonical' solution of the classical dynamical 
Yang-Baxter equation on an arbitrary self-dual Lie algebra.
\end{abstract}

\def\G{{\cal G}}                         %
\def\bR{{\mathbf R}}                     %
\def\End{{\mathrm {End}}}                %
\def\Ad{{\mathrm {Ad}\,}}                %
\def\A{{\cal A}}                         %
\def\F{{\cal F}}                         %
\def\L{{\cal L}}                         %
\def\R{{\cal R}}                         %
\def\B{{\cal B}}                         %

\newpage
\section{Introduction}
\setcounter{equation}{0}

Let $G$ be a connected (real or complex) Lie group 
whose Lie algebra $\G$ is self-dual in the sense of admitting 
a nondegenerate invariant scalar product $\langle\ ,\ \rangle$. 
The corresponding chiral WZNW phase space consists 
of $G$-valued quasiperiodic fields on the real line, 
$g(x+2\pi)=g(x)M$ $(\forall x\in \bR$), 
where $M\in G$ is the `monodromy matrix'. 
After restricting $M$ to some open submanifold $\check G\subset G$ 
(see below), this phase space is equipped~\cite{FG,BFP} 
with a Poisson structure that 
can be symbolically described as follows:  
\be
\kappa \left\{g(x)\stackrel{\otimes}{,} g(y)\right\}
=\left(g(x)\otimes g(y)\right)\Bigl(
\hat r(M) + {1\over 2} {\hat I} \,{\mathrm sign}\,(y-x)
\Bigr), \quad 0< x,y<2\pi, 
\label{1.1}\ee 
where 
$\hat r: \check G \rightarrow \G\otimes \G$, that is 
$\hat r(M)= r^{\alpha\beta}(M) T_\alpha \otimes T_\beta$,
is an {\em antisymmetric} `exchange $r$-matrix', and 
$\hat I= T_\alpha \otimes T^\alpha$
using dual bases $\{ T_\alpha\}$, $\{T^\beta\}$ of $\G$;
$\kappa$ is some constant.
The Jacobi identity of the Poisson bracket (\ref{1.1}) 
is equivalent to  a certain dynamical generalization of the (modified) classical
Yang-Baxter equation on the exchange $r$-matrix, which we call 
the `$G$-CDYBE'. 
If for any function $\psi$ on $G$ we introduce 
\be
({\cal L}_\alpha \psi)(M):= {d\over d t}
 \psi(e^{t T_\alpha} M)\Big\vert_{t=0},
\qquad
 ({\cal R}_\alpha \psi)(M):= {d\over d t}
 \psi(Me^{t T_\alpha} )\Big\vert_{t=0},
\label{1.3}\ee
then the $G$-CDYBE~\cite{BFP} is given as follows:
\be
\big[\hat r_{12},\hat r_{23}\big]
+T^\alpha_1
\big({1\over 2} {\cal D}^+_\alpha  
+ r_\alpha^{\phantom{\alpha} \beta}{\cal D}^-_\beta\big) 
\hat r_{23}
+ \hbox{cycl. perm.}=
-{1\over 4} \hat f.
\label{1.4}\ee
Here ${\cal D}_\alpha^\pm:= {\cal R}_\alpha \pm {\cal L}_\alpha$,
the cyclic permutations act on the three tensorial factors as usual,
and 
$\hat f=f_{\alpha\beta\gamma}T^\alpha\otimes T^\beta\otimes T^\gamma$ 
with the structure constants of $\G$. 
The Poisson structure (\ref{1.1})  arises by inverting~\cite{BFP} 
a (weak) symplectic form on the chiral WZNW phase space,
which is defined~\cite{FG} with the aid of a $2$-form $\rho$ on $\check G$ whose
exterior derivative coincides with the restriction of the 
canonical 3-form on $G$. 
In this context, the restriction to some $\check G\subset G$ 
is necessary since the canonical 3-form on $G$ is closed but not exact.
In the next section we 
provide a self-contained {\em finite-dimensional} characterization 
of the relationship between the 2-form $\rho$ and the exchange 
$r$-matrix $\hat r$.
This relationship has previously been established~\cite{BFP} by indirect arguments
relying on the properties of the {\em infinite-dimensional} 
chiral WZNW phase space.

\section{The 2-form $\rho$ and the exchange $r$-matrix} 
\setcounter{equation}{0}

For any (real or complex) manifold $Q$, denote by
${\cal F}(G, Q)$ the set of (smooth or holomorphic)
maps from $G$ to $Q$, and let $\F(G)$ be the set of (smooth or holomorphic)
scalar functions on $G$.
Let $\bar M \in \F(G,\End(\G))$ be the map
\begin{equation}
\bar M: M\mapsto \Ad M
\qquad
\forall M\in G.
\label{2.1}\end{equation}
With any $X\in \F(G,\G)$ associate the vector field ${\cal R}_X$ on $G$
by the definition
\begin{equation}
(\R_X \psi)(M):= {d\over d t}
 \psi(Me^{t X(M)} )\Big\vert_{t=0},
\qquad
\forall M\in G,\quad \psi\in \F(G).
\label{2.2}\end{equation}
Let $\check G$ be an open domain in $G$ and
choose a function $r\in \F(\check G, \End(\G))$.
Define $r_\pm := r \pm \frac{1}{2} I$ and 
introduce 
$\A\in \F(\check G, \End(\G))$ as
\begin{equation}\A := (\bar M^{-1} \circ r_+ - r_-).
\label{2.3}\end{equation}
With these notations, the $G$-CDYBE (\ref{1.4}) 
can be rewritten as follows:
\begin{equation}
\langle [r \xi, r\eta], \zeta\rangle
- \langle \R_{\A(\xi)} r \eta, \zeta \rangle
+ \mbox{ c.p.}
= -\frac{1}{4} \langle [\xi, \eta],\zeta \rangle
\qquad
\forall \xi, \eta, \zeta \in \G,
\label{2.4}\end{equation}
where the cyclic permutations act on $\xi, \eta, \zeta$.

Consider the canonical 3-form $\Phi$ on $G$, 
\begin{equation}
\Phi(\R_X, \R_Y, \R_Z):= \frac{1}{2} \langle [X,Y], Z\rangle,
\label{2.5}\end{equation}
where $[X,Y](M):= [X(M), Y(M)]$.
Let $q\in \F(\check G, \End(\G))$ be a map for which
$q(M)$ is antisymmetric with respect to $\langle\ ,\ \rangle$ for any
$M\in \check G$.
With such a map $q$ associate a 2-form $\rho$ on $\check G$ by 
\begin{equation}
\rho(\R_X, \R_Y)= \langle X, q Y\rangle
\qquad
\forall X,Y\in \F(\check G, \G).
\label{2.6}\end{equation}

\medskip
\noindent
{\bf Proposition 1.}
{\em Suppose that $d\rho =\Phi$ on an open submanifold $\check G\subset G$ and
\begin{equation}
\det \left( q_+(M) -  q_-(M)\circ  \Ad (M^{-1}) \right)\neq 0
\qquad
\forall M\in \check G,
\label{2.7}\end{equation}
where $q_\pm := q\pm \frac{1}{2} I$. Then
\be
r := {1\over 2}
\left( q_+ -  q_-\circ \bar M^{-1}\right)^{-1}
\circ
\left(q_+ + q_- \circ \bar M^{-1}  \right)
\label{2.8}\ee
solves the $G$-CDYBE (\ref{2.4}) on $\check G$.
}
\medskip

\noindent
{\em Remark.}
In fact, under the assumption (\ref{2.7}) 
formula (\ref{2.8}) gives the unique solution 
of the factorization equation   
\be
q_+ \circ r_-= q_-\circ \bar M^{-1} \circ r_+ 
\label{2.9}\ee
for $r$, and one also has the equivalent formula
\begin{equation}
r_- = - q_- \circ \left( q_- - \bar M\circ q_+\right)^{-1}.
\label{2.10}\end{equation}

\noindent
{\bf Proof of proposition 1.}
We start by noting that, 
as a special case of a standard identity between the Lie and
the exterior derivatives, any 2-form $\rho$ on $G$ (or $\check G$)
satisfies 
\begin{equation}
(d\rho)(\R_X,\R_Y,\R_Z)=\R_X(\rho(\R_Y,\R_Z)) -\rho([\R_X,\R_Y], \R_Z)
+ \mbox{c.p.},
\label{2.11}\end{equation}
where the cyclic permutations act on $X,Y,Z$.
The Lie bracket of the vector fields $\R_X, \R_Y$ is given by
\be
[ \R_X, \R_Y] = \R_{\B(X,Y)}
\quad
\hbox{with}\quad
\B(X,Y)= [X,Y] + \R_X Y - \R_Y X.
\label{2.12}\ee
By using the Leibniz rule to evaluate the first term of (\ref{2.11})
and inserting (\ref{2.12}) into (\ref{2.11}), (\ref{2.6}) implies that 
\begin{equation}
(d\rho)(\R_X,\R_Y,\R_Z) = 
\langle q [X,Y], Z\rangle - \langle (\R_X q) Y,Z\rangle
+\hbox{c.p.}
\label{2.13}\end{equation}
Now taking into account the assumption $d\rho=\Phi$, we obtain
the  identity  
\be
-\langle (\R_X q) Y,Z\rangle 
+\hbox{c.p} = \langle  [X,Y], qZ\rangle +\frac{1}{6}
 \langle [X, Y],Z\rangle
+\hbox{c.p.}
\label{2.14}\ee
$\forall X, Y, Z\in \F(\check G,\G)$.
Then a further important identity is   
\begin{equation}
\langle \R_{\A(\xi)} r \eta, \zeta \rangle = 
\langle (\R_{\A(\xi)} q) \A(\eta), \A(\zeta)\rangle 
- \langle (\bar M \circ r_- - r_+) \xi, [ r_+ \eta, r_+\zeta]\rangle
\label{2.15}\end{equation}
$\forall \xi, \eta, \zeta \in \G$.  
One can derive this by calculating the derivative 
$\R_{\A(\xi)} r$ from the formula (\ref{2.10}), and using many times
(\ref{2.9}), the invariance of $\langle\ ,\ \rangle$, and 
the antisymmetry of $r$ and $q$.
The identities (\ref{2.14}), (\ref{2.15}) allow us to express the 
derivative term in the $G$-CDYBE (\ref{2.4}) in 
terms of non-derivative terms containing $r$.
More precisely, we need one more identity to do this,
\begin{equation}
\langle [\A(\xi), \A(\eta)], q\A(\zeta)\rangle =
\frac{1}{2} \langle [\A(\xi), \A(\eta)], \A(\zeta)\rangle +
 \langle [\A(\xi), \A(\eta)], r_- \zeta\rangle,
\label{2.16}\end{equation}
which follows from the definition of $\A$ 
(\ref{2.3}) and (\ref{2.10}).
With the last three identities in hand, the $G$-CDYBE 
(\ref{2.4}) can be verified straightforwardly  as
\begin{eqnarray}
&&\langle[r\xi, r\eta], \zeta\rangle 
- \langle \R_{\A(\xi)} r \eta, \zeta \rangle +\hbox{c.p.} = 
\langle[r\xi, r\eta], \zeta\rangle 
 - \langle (\R_{\A(\xi)} q) \A(\eta), \A(\zeta)\rangle 
\nonumber\\
&&\phantom{\langle[r\xi, r\eta], \zeta\rangle 
- \langle \R_{\A(\xi)}[r] \eta, \zeta \rangle +\hbox{c.p.}}
\quad + \langle (\bar M \circ r_- - r_+)\xi,
[r_+\eta, r_+\zeta]\rangle +\hbox{c.p.} 
\nonumber\\
&&
\qquad\qquad 
=\langle[r\xi, r\eta], \zeta\rangle
+ \langle (\bar M \circ r_- - r_+)\xi, [r_+\eta, r_+\zeta]\rangle
\nonumber\\
&&
\qquad \qquad\qquad
+\langle [\A(\xi), \A(\eta)], q\A(\zeta)\rangle 
+\frac{1}{6} \langle [\A(\xi), \A(\eta)], \A(\zeta)\rangle +\hbox{c.p.}
\nonumber\\
&& 
\qquad\qquad
=  \langle[r\xi, r\eta], \zeta\rangle
+ \langle [\bar M^{-1} \circ r_+\xi, \bar M^{-1} \circ r_+\eta], 
r_- \zeta \rangle
\nonumber\\
&&
\qquad\qquad\qquad 
-  \langle [\bar M^{-1} \circ r_+\xi, \bar M^{-1} \circ r_+\eta], 
\bar M^{-1}\circ r_+ \zeta \rangle
\nonumber\\
&&
\qquad\qquad\qquad
+\langle [\A(\xi), \A(\eta)], r_-\zeta\rangle
+ \frac{2}{3} \langle [\A(\xi), \A(\eta)], \A(\zeta)\rangle +\hbox{c.p.}
\nonumber\\
&&
\qquad\qquad
=  \langle[r\xi, r\eta], \zeta\rangle
- \frac{1}{3} \langle [\bar M^{-1} \circ r_+\xi, \bar M^{-1} 
\circ r_+\eta], \bar M^{-1}\circ r_+ \zeta \rangle
\nonumber\\
&&
\qquad\qquad\qquad
+ \frac{1}{3}\langle[r_-\xi, r_-\eta], r_-\zeta\rangle +\hbox{c.p.}
= -\frac{1}{4}\langle [\xi, \eta], \zeta \rangle,
\label{2.17}\end{eqnarray}
which proves the proposition.
{\em Q.E.D.}

\medskip
\noindent
{\bf Proposition 2.}
{\em Suppose that $r\in \F(\check G, \End(\G))$ solves the 
$G$-CDYBE (\ref{2.4}) on an open submanifold $\check G\subset G$ and
\begin{equation}
\det \left( \Ad (M^{-1})\circ r_+(M) - r_-(M) \right)\neq 0
\qquad
\forall M\in \check G.
\label{2.18}\end{equation}
Then the $2$-form $\rho$ defined by (\ref{2.6}) with 
\be
q := {1\over 2}
\left( \bar M^{-1} \circ r_+ + r_-  \right)
\circ
\left( \bar M^{-1} \circ r_+ - r_- \right)^{-1}
\label{2.19}\ee
satisfies $d\rho =\Phi$ on $\check G$.
}
\medskip

\noindent
{\bf Proof.}
Consider the equivalent formula of the $q$
\begin{equation}
q_- = r_- \circ \left( \bar M ^{-1}\circ r_+ - r_-\right)^{-1}
= r_- \circ \A ^{-1},
\label{2.20}\end{equation}
where $\A$ (\ref{2.3}) is invertible due to (\ref{2.18}).
By taking the derivative of this equation one can show that
relation (\ref{2.15}) is valid in this case as well.
By combining (\ref{2.15}) with (\ref{2.4}) we obtain that 
\begin{eqnarray}
&&\langle (\R_{\A(\xi)} q) \A(\eta), \A(\zeta) \rangle + \hbox{c.p.} 
=\nonumber\\
&&= \langle [r\xi, r\eta], \zeta \rangle
+ \frac{1}{12} \langle [\xi, \eta], \zeta\rangle 
- \langle (r_+ - \bar M\circ r_-)\xi, [r_+\eta, r_+\zeta]\rangle
+\hbox{c.p.}\nonumber\\
&&=\langle q [\A(\xi), \A(\eta)], \A(\zeta) \rangle
- \frac{1}{6} \langle [\A(\xi), \A(\eta)], \A(\zeta) \rangle
+ \hbox{c.p.}
\label{2.22}\end{eqnarray}
$\forall \xi, \eta, \zeta \in \G$.
The second equality here is derived
by using several times (\ref{2.20}) and the 
corresponding equation for $q_+$, i.e.,
$q_+={\bar M}^{-1} \circ r_+ \circ \A^{-1}$.
By (\ref{2.13}) and (\ref{2.5}), 
this allows us to conclude that 
\be
(d\rho)(\R_{\A(\xi)},\R_{\A(\eta)},\R_{\A(\zeta)}) = 
\Phi(\R_{\A(\xi)}, \R_{\A(\eta)}, \R_{\A(\zeta)}). 
\label{2.23}\ee
Since the correspondence $\xi \mapsto \A(\xi)$
is invertible due to (\ref{2.18}), this implies that 
$d\rho =\Phi$ holds.  {\em Q.E.D.}

\section{Recovering a canonical solution of the CDYBE}
\setcounter{equation}{0}

Let us now suppose that $\check G$ is diffeomorphic to a domain 
$\check \G\subset \G$ by the exponential parametrization, whereby
we write $\check G\ni M =e^{\omega}$ with $\omega\in \check \G$.
Then choose~\cite{BFP} the 2-form $\rho$ on $\check G \simeq \check \G$ 
to be
\be
\rho(\omega):= -\frac{1}{2} \int_0^1 dx 
\langle d \omega \stackrel{\wedge}{,}de^{x\omega} \, e^{-x\omega}\rangle.
\label{3.1}\ee
It is not difficult to check that  $d\rho = \Phi$ holds
and one can also calculate the explicit form of the operator $q(\omega)$ 
that corresponds to $\rho$ in (\ref{3.1}) by means of (\ref{2.6}).
Let ${\cal Q}$ and ${\cal R}$ denote the complex meromorphic functions 
\be
{\cal Q}(z)= \frac{2z + e^{-z} - e^z } {2 (e^z -1) (1-e^{-z})},
\qquad 
{\cal R}(z):= \frac{1}{2} \coth \frac{z}{2} - \frac{1}{z},
\label{3.2}\ee 
which are regular around $z=0$.   
In fact, (\ref{3.1}) implies the formula
\be
q(\omega)= {\cal Q}(\mathrm{ad}\, \omega),
\label{3.3}\ee
where the right hand side is defined by means of the 
power series expansion of the function ${\cal Q}$ around $z=0$
if $\omega$ is near to the origin.  
By inserting this into equation (\ref{2.9}) that 
determines $r$ in terms of $q$,
one finds that the exchange $r$-matrix associated with the
$2$-form $\rho$ in (\ref{3.1}) is given by  
\be
r(\omega) = {\cal R}(\mathrm{ad}\, \omega).
\label{3.5}\ee

Consider  the holomorphic complex function 
\be
h: z \mapsto  \frac{e^z -1}{z},
\ee
and recall that for a curve
$t \mapsto A(t)$
of finite-dimensional linear operators  
\begin{equation}
\frac{ d e^{\pm A(t)}}{dt}= \pm e^{\pm A(t)} 
h(\mp {\mathrm{ad}}_{A(t)}) (\dot{A}(t)), 
\qquad
\dot{A}(t):= \frac{ dA(t)}{ dt}.
\label{3.7}\end{equation}
The right hand side of this equation is defined by means  of 
the  Taylor expansion of $h(z)$ around $z=0$, and    
$\mathrm{ad}_{A(t)}(\dot{A}(t))= [A(t),\dot{A}(t)]$. 
By using relation (\ref{3.7})  it is not difficult to derive (\ref{3.3}).
Relation (\ref{3.7}) also implies that the $r$-matrix in (\ref{3.5}) 
satisfies the following identity:
\be
\big({1\over 2} {\cal D}^+_\alpha  
+ r_\alpha^{\phantom{\alpha} \beta}{\cal D}^-_\beta\big) 
= \frac{\partial} {\partial \omega^\alpha},
\label{3.8}\ee
where we use that 
$M=e^\omega$ and 
${\cal D}_\alpha^\pm = ({\cal R}_\alpha \pm {\cal L}_\alpha)$ 
are given by (\ref{1.3}).
Incidentally, requiring (\ref{3.8}) for 
{\em the ansatz} $r(\omega)= \R(\mathrm{ad}\, \omega)$ 
with some odd function $\R(z)$ that is holomorphic around the origin
leads uniquely to the function $\R(z)$ in (\ref{3.2}).

As a consequence of proposition 1, we know that 
$r(\omega)= {\cal R}(\mathrm{ad}\, \omega)$ with (\ref{3.2})
satisfies the $G$-CDYBE (\ref{1.4}).
We now notice from (\ref{3.8})  
that {\em for this particular $r$-matrix}  (\ref{1.4}) 
becomes the standard CDYBE~\cite{EV} on the Lie algebra $\G$: 
\be
\big[\hat r_{12},\hat r_{23}\big]
+T^\alpha_1
\frac{\partial} {\partial \omega^\alpha} \hat r_{23}
+ \hbox{cycl. perm.}=
-{1\over 4} \hat f.
\label{3.9}\ee
We call the $r$-matrix in (\ref{3.5}) the `canonical' solution of the 
CDYBE (\ref{3.9}) on $\G$
since many solutions of the CDYBE with respect to 
subalgebras ${\cal H}\subset {\cal G}$ can be 
derived from it by applying Dirac reduction~\cite{FGP1} 
to the dynamical variable $\omega$. 
The above proof of the statement that the canonical $r$-matrix solves 
(\ref{3.9}) was extracted from the study of the WZNW model, 
where it arose in a natural manner~\cite{BFP}.
In this context, (\ref{3.8}) is in fact equivalent to  having 
the Poisson brackets $\kappa \{ g(x), \omega_\alpha\} =g(x) T_\alpha$,
which means that the logarithm of the monodromy matrix serves as
the infinitesimal generator (momentum map)
for a classical $\G$-symmetry on the chiral WZNW phase space.
For other applications of the canonical $r$-matrix and its reductions, 
and for different verifications of the fact that it solves the CDYBE, 
we refer to the literature~\cite{EV,AM,FP2}.

\bigskip
\bigskip
\noindent
{\bf Acknowledgements.}
This investigation was supported in part by the Hungarian 
Scientific Research Fund (OTKA) under T034170 and M028418. 

\newpage
\renewcommand{\thesection}{\Alph{section}}
\setcounter{section}{0} 

\section{Some calculations}
\setcounter{equation}{0}
\renewcommand{\theequation}{A.\arabic{equation}}

For convenience, we here sketch the derivation of 
some of the statements mentioned in the text.

First, let us show that $d\rho = \Phi$ holds for $\rho$ 
in (\ref{3.1}). 
By using the exponential parametrization,  we 
introduce the $\G$-valued 
$1$-form $\theta_x:= (d e^{x\omega}) e^{-x \omega}$ for 
any $0\leq x\leq 1$. 
Plainly, $\Phi$ (\ref{2.5}) can be written as  
\be
\Phi = \frac{1}{2} \langle \theta_1, [ \theta_1, \theta_1]\rangle
= \frac{1}{2} \int_0^1 dx\, \frac{d}{dx}
\langle \theta_x, [ \theta_x, \theta_x]\rangle
\quad \hbox{on}\quad \check G=\exp(\check \G). 
\label{a.1}\ee
By noting that $\frac{d \theta_x}{dx}=[\omega, \theta_x] +d\omega$,
we obtain 
\bea
&&\frac{d}{dx}
\langle \theta_x, [ \theta_x, \theta_x]\rangle
=\langle \langle [\omega, \theta_x], [\theta_x, \theta_x]\rangle 
+ \langle \theta_x, [[\omega, \theta_x], \theta_x]\rangle 
+\langle \theta_x, [\theta_x,[\omega, \theta_x]]\rangle\nonumber\\
&&\qquad\qquad + \langle d\omega, [\theta_x, \theta_x]\rangle 
 + \langle \theta_x, [d\omega, \theta_x]\rangle
+ \langle \theta_x, [\theta_x, d\omega]\rangle
 = \langle d \omega \stackrel{\wedge}{,} 
[\theta_x, \theta_x]\rangle.
\label{a.3}\eea
In the second equality we used the invariance 
of $\langle\ ,\ \rangle$ and the usual rules of calculation
for (Lie algebra valued) differential forms\footnote{For example, 
if $\beta$, $\gamma$ are $\G$-valued 1-forms and $A$, $B$ are vector fields,
then $[\beta,\gamma]$ is the $\G$-valued 2-form 
$[\beta,\gamma](A,B)=[\beta(A), \gamma(B)]$;
$\langle \beta \stackrel{\wedge}{,} \gamma \rangle (A,B)=
\langle \beta(A), \gamma(B)\rangle - \langle \beta(B), \gamma(A)\rangle$; 
and analogously for for $k$-forms.}. 
It follows from (\ref{3.1}) that  
\be
d\rho= \frac{1}{2} \int_0^1 dx\, \langle d \omega \stackrel{\wedge}{,} 
[\theta_x, \theta_x]\rangle,
\label{a.4}\ee
and therefore $d\rho = \Phi$ holds indeed on $\exp(\check \G)$. 

Next, let us describe how (\ref{3.1}) leads to formula (\ref{3.3})
of the operator $q(\omega)$ by means of (\ref{2.6}). For this,
we need the complex functions
\be
h_1(z)=h(z)=\frac{e^z -1}{z},\qquad
h_2(z)= h(-z),
\qquad 
h_i^{-1}(z)=\frac{1}{h_i(z)}.
\label{A.1}\ee
The well known relation (\ref{3.7}) implies that
\be
\alpha(\omega):= -\frac{1}{2}\int_0^1 dx\, 
(de^{x\omega})  e^{-x\omega} =
-\frac{1}{2}\int_0^1 dx\, 
h(x\,{\mathrm{ad}}\,\omega) (x\,d\omega) =
f({\mathrm{ad}}\,\omega) (d\omega)
\label{A.2}\ee
with the holomorphic function
\be
f(z)= \frac{1}{2z} +\frac{1-e^z}{2z^2}.
\label{A.3}\ee
The third equality in (\ref{A.2}) follows by
termwise integration of the series of $x h(x \,{\mathrm{ad}}\,\omega)$.  
Thus $\rho(\omega)$ in (\ref{3.1}) can be written as
\be
\rho(\omega)= 
\langle d \omega \stackrel{\wedge}{,} \alpha(\omega)\rangle
=\langle d \omega \stackrel{\wedge}{,} 
f({\mathrm{ad}}\,\omega)(d\omega)\rangle
=\langle d \omega , 
F({\mathrm{ad}}\,\omega)(d\omega)\rangle
\label{A.4}\ee
with
\be
F(z)= f(z)-f(-z).
\label{A.5}\ee
As another consequence of (\ref{3.7}), we have
\be
d\omega = h_2^{-1}({\mathrm{ad}}\,\omega) (M^{-1} dM).
\ee
By substituting this into (\ref{A.4}) and using the invariance of
$\langle\ ,\ \rangle$, we obtain  
\be
\rho(\omega)=\langle M^{-1}d M , 
{\cal Q}({\mathrm{ad}}\,\omega)(M^{-1}dM)\rangle,
\label{A.6}\ee
where
\be
{\cal Q}(z)= \frac{1}{h_1(z)} F(z) \frac{1}{h_2(z)}= 
\frac{2z + e^{-z} - e^z } {2 (e^z -1) (1-e^{-z})}.
\label{A.7}\ee
By means of (\ref{2.6}), this is clearly equivalent to the
equality claimed in (\ref{3.3}). 

It is readily verified that the functions ${\cal Q}(z)$ and 
${\cal R}(z)$ in (\ref{3.2}) enjoy the identity 
\be
\left({\cal Q}(z) + \frac{1}{2}\right)\left({\cal R}(z) -\frac{1}{2}\right)= 
\left({\cal Q}(z) - \frac{1}{2}\right)e^{-z}\left({\cal R}(z) +\frac{1}{2}\right), 
\label{E.1}\ee
which implies the factorization equation (\ref{2.9}) for 
the corresponding operators $q(\omega)={\cal Q}({\mathrm{ad}\,} \omega)$ and
$r(\omega)={\cal R}({\mathrm{ad}}\,\omega)$. 
Let us also remark that (\ref{3.8}) is guaranteed by the 
following identity satisfied by the function ${\cal R}(z)$: 
\be
\frac{1}{2}( h^{-1}_1 + h^{-1}_2 ) + {\cal R}( h^{-1}_1 - h^{-1}_2)=1.
\label{E.2}\ee
To see that this implies (\ref{3.8}), one needs to express 
${\cal L}_\alpha$, ${\cal R}_\alpha$ with the aid
of the relation (\ref{3.7}) as 
\be
{\cal L}_\alpha = \left(h_2^{-1}({\mathrm{ad}}\,\omega)\right)_{\alpha}
^{\phantom{\alpha}\beta}\frac{\partial}{\partial \omega^\beta},
\qquad
{\cal R}_\alpha = \left(h_1^{-1}({\mathrm{ad}}\,\omega)\right)_{\alpha}
^{\phantom{\alpha}\beta}
\frac{\partial}{\partial \omega^\beta} .
\label{E.3}\ee 
Here the functions $h_1$, $h_2$ are given in (\ref{A.1}),
and in particular $h_1^{-1}(z)- h_2^{-1}(z)=-z$.
It is thus clear that (\ref{E.2}) determines ${\cal R}(z)$ uniquely,
and hence (\ref{3.8}) leads uniquely to the canonical $r$-matrix
upon the ansatz $r(\omega)={\cal R}({\mathrm{ad}}\,\omega)$, 
as mentioned in section 3.

\end{document}